\documentclass[12pt]{article}
\usepackage[]{epsfig}
\usepackage[centertags]{amsmath}
\usepackage{amsfonts}
\usepackage{amssymb}
\usepackage[english]{babel}
\usepackage{amsmath, amsthm, slashed,url}

\begin{document}

\title{\bf Supersymmetric soliton solution in (1+1)-dimensional Ultracold Quantum Gases }
\author{
Lucas Sourrouille$^a$
\\
{\normalsize \it $^a$Departamento de F\'\i sica, FCEyN, Universidad
de Buenos Aires}\\ {\normalsize\it Pab.1, Ciudad Universitaria,
1428, Ciudad de Buenos Aires, Argentina}
\\
{\footnotesize  sourrou@df.uba.ar} } \maketitle

\abstract{We obtain by dimensional reduction a $(1+1)$ supersymmetric system introduced in the description of ultracold quantum 
gases. The correct supercharges are identified and their algebra is constructed. 
Finally novel static self-dual solutions emerge, satisfying a Liouville type  differential equation. }


\newpage


\vspace{1cm}
\section{Introduction}

Since its birth in the early $1970$'s in the context of high energy physics and mathematical physics, supersymmetry has found a
growing number of applications strongly influencing many areas both in experimental and theoretical physics.

Originally proposed as a graded extension of the Poincar\'{e} algebra\cite{R}, it was soon recognized that it can also be 
considered in the systems that exhibit Galilean invariance and this lead to the construction of a graded super-Galilean 
algebra\cite{Puza} in $d=3+1$ space time dimensions. Constructing the supersymmetric extension of the Galilean invariant $d=2+1$ 
Jackiw-Pi model\cite{JP}, Leblanc et al \cite{Lozano} discovered the existence of $2$ graded superalgebras and related this to the 
possibility of finding $BPS$ equations in the bosonic sector. After that, there were several developments related to Galilean 
supersymmetry in diverse contexts\cite{horv}.

In $(1+1)$ dimension Galilean supersymmetry was considered to study ultra cold quantum gases\cite{Van2,Van1,my}.
Ultracold quantum gases not only are interesting by the physics that they describe, but also are a useful tool in the modeling of 
others branches in physics\cite{Khawaja01}. One interesting example of this modeling was considered in Ref.\cite{Van2,Van1}. 
There, the authors propose that by combining a vortex line in a one-dimensional optical lattice with a fermionic gas bound to the 
vortex core, it is possible to tune the laser parameters such that a nonrelativistic supersymmetric string is created. This could 
allow to test experimentally several aspects of superstring and supersymmetry theory.

From the theoretical point of view, the model that describe that proposal presents a supersymmetric structure. Despite the theory
has interactions, the authors only found the generators of the super-Galilei algebra for the free theory. This fact constrast with 
a basic feature of supersymmetry theory, where the full Hamiltonian is generated by the supercharge algebra. Later, in 
ref.\cite{my} was shown the existence of supersymmetry charge whose algebra generate the Hamiltonian with quartic interactions. 
Nevertheless a supersymmetry transformation associated to a charge be able to generate the full Hamiltonian of the theory was not 
found. In this paper we will show the existence of a supersymmetry related to a charge generating the full Hamiltonian of the 
theory presents on Ref.\cite{Van2,Van1}. Also we will construct the complete supersymmetry algebra. In order to discus this 
aspects we will present the $(1+1)$ model study in Ref.\cite{Van1} as the dimensional reduction of a Maxwell-Chern-Simons model 
proposed by Manton\cite{Manton1}. This has no influence in the derivation of the correct supersymmetry generators but leads us to 
the presences of interesting solitonic equations in the system. For the bosonic sector these equations reduce to a Liouville type 
differential equation. We analyze this case and construct the solution for this type of Liouville equation.

\section{The Model}

Let us start by considering some features of the model proposed by Manton. This model is governed by a (2+1)-dimensional action consisting on a mixture from the standard Landau-Ginzburg and the Chern-Simons model, where the matter is represented by a complex scalar field $\phi(x)$,

\begin{eqnarray}
S_{(2+1)} = \int d^3 x \Big(-\frac{1}{2} B^2 + i\gamma ( \phi^\dagger \partial_t \phi + i A_0 |\phi|^2 )
- \frac{1}{2m} ( D_i\phi )^\dagger  D_i\phi +\nonumber \\
\kappa (A_0 B + A_2 \partial_0 A_1) + \gamma A_0 + \lambda ( |\phi|^2 -1)^2 -A_i J_i^T \Big)
\label{Ac}
\end{eqnarray}

Here $\gamma$, $\kappa$ and $\lambda$ are real constants, $D_{\mu}= \partial_{\mu} + iA_{\mu}$ $(\mu =0,1,2)$ is the covariant derivative and $B =\partial_{1}A_{2}- \partial_{1}A_{2}$ the magnetic field. The term $\gamma A_0$ is related to the possibility of a condensate in the ground state\cite{ByH} and $J_i^T$ is a constant transport current.
It was show by Manton in Ref.\cite{Manton1} that this theory presents Galilean invariance with the requirement that the transport current transforms as $J_i^T \rightarrow J_i^T + \gamma v_i$ under a boost. With this consideration, we can choose a frame where $J_i^T =0$. The field equations in this frame takes the form

\begin{eqnarray}
i\gamma D_0 \phi = -\frac{1}{2} D_i D_i \phi -2\lambda (|\phi|^2 -1) \phi
\nonumber \\[3mm]
\epsilon_{i j} \partial_j B = J_i + \kappa \epsilon_{i j} E_j
\nonumber \\[3mm]
\kappa B =  \gamma (|\phi|^2 -1)
\label{Eq}
\end{eqnarray}

where $E_i =\partial_{i}A_{0}- \partial_{0}A_{i}$ is the electric field and  $J_i$ is the supercurrent defined by

\begin{eqnarray}
J_i= -\frac{i}{2}  \Big( \phi^\dagger  D_i \phi -\phi (D_i \phi)^\dagger \Big)
\end{eqnarray}

The first equation of this system is the non-linear Schr\"{o}dinger equation.The second is the Amp\`{e}re's law in two dimensions.The last equation is the Chern-Simons version of the Gauss law, which here takes a different form from that presents in the Jackiw-Pi model\cite{JP}. The energy of the system for static field configurations reads as

\begin{eqnarray}
E = \int d^3 x \Big(\frac{1}{2} B^2
+ \frac{1}{2m} ( D_i\phi )^\dagger  D_i\phi -\lambda ( |\phi|^2 -1)^2 \Big)
\label{H1}
\end{eqnarray}

For finiteness we require that the energy vanishes asymptotically. This fixes the asymptotic behavior of the fields

\begin{eqnarray}
\lim_{r \to \infty} \phi(x) =  e^{i\alpha(\phi)}
\,,
\;\;\;\;\;\
\lim_{r \to \infty} A_i = \partial_i \alpha
\end{eqnarray}

where $\alpha$ is common phase angle. With these conditions the magnetic flux reads

\begin{eqnarray}
\Phi =& \int \,\,d^2 x B = \oint_{|x|=\infty} \,\, A_i dx^i  = 2\pi N
\label{}
\end{eqnarray}
where $N$ is a topological invariant which takes only integer values.
Following to Hassa\"{i}ne et al.\cite{HHY}, we can rewrite the expression (\ref{H1}) as

\begin{eqnarray}
 E= \int d^3 x \Big( \frac{1}{2m} |(D_1 \pm iD_2)\phi|^2  +(\mp \frac{\gamma }{2\kappa m} +\frac{\gamma^2}{2\kappa^2} -\lambda) ( |\phi|^2 -1)^2 \mp \frac{1}{2m} B\Big)
\label{}
\end{eqnarray}

where we have used the Gauss law of the equations (\ref{Eq}) and the identity $|D_i \phi|^2 = |( D_1 \pm iD_2)\phi|^2 \mp B|\phi|^2 \pm \epsilon^{ij} \partial_i J_j$. For $\lambda =  \mp \frac{\gamma }{2\kappa m} +\frac{\gamma^2}{2\kappa^2}$ the
potential terms cancel, and we see that the energy is bounded below by a multiple of the magnitude
of the magnetic flux (for positive flux we choose the lower signs, and for negative flux we choose
the upper signs):

\begin{eqnarray}
E \geq \frac{1}{2m} |\Phi|
\label{Emas}
\end{eqnarray}

So this bound is saturated by fields obeying the self-duality equations

\begin{eqnarray}
(D_1 \pm iD_2)\phi =0
\nonumber \\[3mm]
\kappa B =  \gamma (|\phi|^2 -1)
\label{}
\end{eqnarray}
Motivated by these results and the previous works on the (1+1)-dimensional supersymmetry in ultracold quantum gases\cite{Van1,Van2,my} we are interested here on the dimensional reduction of the supersymmetric extension of the model (\ref{Ac}). Such extension can be carried out by considering the inclusion of non-relativistic (down-spinor) fermion $\psi$\cite{Lozano}:

\begin{eqnarray}
S_{(2+1)} = \int d^3 x \Big(-\frac{1}{2} B^2 + i\gamma ( \phi^\dagger  \partial_t \phi + \psi^\dagger  \partial_t \psi + i A_0 [ |\phi|^2 + |\psi|^2] )
- \frac{1}{2m} ( D_i\phi )^\dagger  D_i\phi -\nonumber \\
\frac{1}{2m} ( D_i\psi )^\dagger  D_i\psi
+ \kappa (A_0 B + A_2 \partial_0 A_1) + \gamma A_0  -\frac{1}{2m} \psi^\dagger B\psi +  \lambda_1 ( |\phi|^2 -1)^2 + \lambda_2 ( |\phi|^2 -1)|\psi|^2\Big)
\label{Ac1}
\end{eqnarray}
where the coupling constants are given by

\begin{eqnarray}
\lambda_1=\frac{\gamma}{2m\kappa} + \frac{\gamma^2}{2\kappa^2}\,, \;\;\;\ \lambda_2 =\frac{3\gamma}{2m\kappa} + \frac{\gamma^2}{\kappa^2}
\label{cop1}
\end{eqnarray}

and we have include a Pauli term for the fermion corresponding to a down-spinor. This action is invariant under the following supersymmetry transformation

\begin{eqnarray}
\delta_{1} \phi &=& \sqrt{2m} \eta_{1}^\dagger  \psi  \;, \;\;\;
\;\;\;
\delta_{1} \psi= - \sqrt{2m} \eta_{1} \phi\;, \nonumber \\
\delta_{1} {\bf A} &=& 0 \;, \;\;\; \;\;\;
\qquad\quad\ \delta_{1} A^{0}=
\frac{1}{\sqrt{2m}c \kappa} ( \eta_{1} \phi \psi^\dagger  -
\eta^\dagger _{1} \psi \phi^\dagger  ) \;. \label{var1}
\end{eqnarray}

if the coupling constants satisfy

\begin{eqnarray}
\frac{\gamma}{2m\kappa} + 2\lambda_1 -\lambda_2 = 0
\end{eqnarray}

Where $\eta_1$ appearing in (\ref{var1}) is a complex Grassmann variable.
In order to analyze the lineal problem\cite{my,JPR}, it is natural to consider a dimensional reduction of the action (\ref{Ac1})
by suppressing dependence on the second spacial coordinate, renaming $A_y$ as $B$. Then, the action (\ref{Ac1}) becomes

\begin{eqnarray}
S_{(1+1)} = \int d^2 x \Big(-\frac{1}{2} (\partial_x B)^2 + i\gamma ( \phi^\dagger  \partial_t \phi + \psi^\dagger  \partial_t \psi + i A_0 \rho )
- \frac{1}{2m} ( D_x \phi )^\dagger  D_x \phi -\frac{1}{2m}B^2 \rho -\nonumber \\
\frac{1}{2m} ( D_x \psi )^\dagger  D_x \psi
+ \kappa (A_0 \partial_x B + B \partial_0 A_1) + \gamma A_0 + \lambda_1 ( \rho_b -1)^2 + \lambda_2 ( \rho_b -1)\rho_f \Big)
\label{Ac2}
\end{eqnarray}

Where we have introduced the matter densities,
\begin{eqnarray}
\rho_b=|\phi|^2 \,,\,\,\,\;\;\;
\rho_f=|\psi|^2 \,,\,\,\,\,\;\;\; \rho=\rho_b+\rho_f
\end{eqnarray}

The Gauss law constraint for this action is

\begin{eqnarray}
\partial_x B= \frac{\gamma}{\kappa} (\rho -1)
\label{G2}
\end{eqnarray}
Note that this constraint has an additional constant term from that appearing on Ref.(\cite{my,JPR}).
The equation can be solved as

\begin{eqnarray} B(x)=\frac{\gamma}{2 \kappa} \int dz \epsilon(x-z) (\rho(z)-1)
\label{gauss} \end{eqnarray}

Using these expressions for the magnetic field and its derivative, the action (\ref{Ac2}) takes the form

\begin{eqnarray}
S_{(1+1)} = \int d^2 x \Big( i\gamma ( \phi^\dagger  \partial_t \phi + \psi^\dagger  \partial_t \psi )
- \frac{1}{2m} ( D_x \phi )^\dagger  D_x \phi - \frac{1}{2m} ( D_x \psi )^\dagger  D_x \psi -\frac{1}{2m}B^2 \rho +\nonumber \\
\kappa \Big( \frac{\gamma}{2 \kappa} \int dz \epsilon(x-z) (\rho(z)-1) \Big)\partial_0 A_1 +  \lambda^{'}_1 ( \rho_b -1)^2 + \lambda^{'}_2 ( \rho_b -1)\rho_f \Big)
\label{Ac3}
\end{eqnarray}

Here the action contains new coupling constants defined as

\begin{eqnarray}
\lambda^{'}_1 =  \lambda_1 -\frac{\gamma^{2}}{2\kappa^{2}} \,,\,\,\,\;\;\;
\lambda^{'}_2 =  \lambda_2 -\frac{\gamma^{2}}{\kappa^{2}} - \frac{\gamma}{2m\kappa}
\label{newc}
\end{eqnarray}

Following the method proposed in  Ref.\cite{JPR}, the gauge field $A_x$ may be eliminate from  the action (\ref{Ac3}) via a 
gauge transformation.
Indeed, after transforming the matter fields as

\begin{eqnarray}
\phi(x)\rightarrow e^{-i\alpha(x)} \phi(x)\,, \,\,\,\,\,
\psi(x)\rightarrow e^{-i\alpha(x)} \psi(x)
\end{eqnarray}

with

\begin{eqnarray}
\alpha(x)=\frac{1}{2}\int dz \epsilon(x-z) A_x(z)
\label{alpha}
\end{eqnarray}

the action can be written simply as

\begin{eqnarray}
S_{(1+1)} = \int d^2 x & \Big(& i\gamma ( \phi^\dagger  \partial_t \phi + \psi^\dagger  \partial_t \psi )
- \frac{1}{2m} ( \partial_x \phi )^\dagger  \partial_x \phi -\frac{1}{2m} ( \partial_x \psi )^\dagger  \partial_x \psi +
\nonumber \\  & &\lambda^{'}_1 ( \rho_b -1)^2 + \lambda^{'}_2 ( \rho_b -1)\rho_f -\frac{1}{2m}B^2 \rho  \Big)
\label{Ac4}
\end{eqnarray}

The last term appearing in this action is a constant of motion,

\begin{eqnarray}
 & &\frac{1}{2m} \int dx_1 B^2(x_1) \rho(x_1)=
\frac{\gamma^2}{8m\kappa^2} \int dx_1 dx_2 dx_3  \epsilon(x_1-x_2) \epsilon(x_1-x_3) (\rho(x_1)-1)\nonumber \\ & &
 (\rho(x_2)-1) \rho(x_3) =
\frac{\gamma^2}{24m \kappa^2} \int dx_1 dx_2 dx_3
(\rho(x_1)-1) (\rho(x_2)-1) \rho(x_3) = \nonumber \\ & &
\frac{ (N^3 -2N^2+N)\gamma^2}{24m \kappa^2}
\end{eqnarray}
where  $N=\int dx \rho(x)$,  and use has been made of the identity

\begin{eqnarray}
\epsilon(x_1-x_2) \epsilon(x_1-x_3) + \epsilon(x_2-x_3)
\epsilon(x_2-x_1) + \epsilon(x_3-x_1) \epsilon(x_3-x_2) = 1\,.
\end{eqnarray}
Thus,  dropping this term, the action can be written as

\begin{eqnarray}
S_{(1+1)} = \int d^2 x & \Big(& i\gamma ( \phi^\dagger  \partial_t \phi + \psi^\dagger  \partial_t \psi )
- \frac{1}{2m} ( \partial_x \phi )^\dagger  \partial_x \phi -\frac{1}{2m} ( \partial_x \psi )^\dagger  \partial_x \psi +
\nonumber \\  & &\lambda^{'}_1 ( \rho_b -1)^2 + \lambda^{'}_2 ( \rho_b -1)\rho_f   \Big)
\label{Ac5}
\end{eqnarray}

This is the model study in Ref.\cite{Van1,Van2}, in the context of description of ultracold gases dynamics. 
The action (\ref{Ac5}) is similar to the action derived by dimensional reduction of the Jackiw-Pi model\cite{my}. The difference 
is that our model also include the chemical potential terms.

Let us now consider the possible supersymmetry transformations that leave unchange the action (\ref{Ac5}). 
One obvious supersymmetry of the system (\ref{Ac5}) takes place when bosons and fermions are interchanged according to

\begin{eqnarray}
\delta_{1} \phi &=& \sqrt{2m} \eta_{1}^\dagger  \psi \;, \;\;\;
\;\;\; \delta_{1} \psi= - \sqrt{2m} \eta_{1} \phi\;,
\label{t1}
\end{eqnarray}

and the coupling constants are related by

\begin{eqnarray}
\lambda^{'}_2 = 2\lambda^{'}_1
\label{cc1}
\end{eqnarray}

It is interesting to note that the transformation (\ref{t1}) is also a supersymmetry of the model study in Ref.\cite{my}.

The fact that is less evident is the existence of a second supersymmetry. In discussing this supersymmetry, note that the the 
action  (\ref{Ac5}) is invariant under the following transformation

\begin{equation}
\begin{array}{ll}
\delta_{2} \phi = \frac{i}{\sqrt{2m}} \eta_{2}^\dagger( \partial_x
\psi - B_1\psi)  + i \delta_2 \alpha \; \phi\; ,  \qquad &\delta_{2} \psi = -
\frac{i}{\sqrt{2m}} \eta_{2} (\partial_x \phi+B_1\phi) + i \delta_2 \alpha \; \psi\,,
 \label{t2}
\end{array}\end{equation}

provided that

\begin{eqnarray}
\lambda^{'}_1=\frac{\gamma}{2m\kappa}\,, \;\;\;\ \lambda^{'}_2 = 2\lambda^{'}_1
\label{cc2}
\end{eqnarray}

and
\begin{eqnarray}
 B_1(x)=\frac{\gamma}{2 \kappa} \int dz \epsilon(x-z) \rho(z)
\nonumber \\[3mm]
\delta_2 \alpha = \frac{-\gamma}{2\kappa \sqrt{2m}} \int dz \epsilon(x-z) \Big(\eta_{2} \phi \psi^\dagger - \eta_{2}^\dagger \psi \phi^\dagger \Big)
\end{eqnarray}

Notice that combining  equations (\ref{newc}) and (\ref{cc2}) we obtain Eq.(\ref{cop1}).
As the transformation  (\ref{t1}), the expression (\ref{t2}) is also a supersymmetry of model explored in Ref.\cite{my}. Another interesting fact is that the condition for coupling constants imposed in (\ref{cc2}) is a particular case of the equation (\ref{cc1}). This implies that the following combination of the precedent two supersymmetries

\begin{eqnarray}
\delta_{2} \phi = \frac{i}{\sqrt{2m}} \eta_{2}^\dagger( \partial_x
\psi - B_1\psi)  + i \delta_2 \alpha \; \phi\; -\frac{i\eta_{2}^\dagger\sqrt{\gamma}}{\sqrt{\kappa m}} \psi ,
\nonumber \\[3mm]
\delta_{2} \psi = -
\frac{i}{\sqrt{2m}} \eta_{2} (\partial_x \phi+B_1\phi) + i \delta_2 \alpha \; \psi\; +\frac{i \eta_{2}\sqrt{\gamma}}{\sqrt{\kappa m}} \phi
\label{t3}
\end{eqnarray}
is also a supersymmetry of the accion (\ref{Ac5}) if the condition (\ref{cc2}) is kept.

\section{The supersymmetry algebra}

In this section we shall study the algebra of the generators associate to the transformations (\ref{t1}) and (\ref{t3}).

In the representation of the supersymmetry algebra the generator associated to the supersymmetry (\ref{t1}) can easily be fond to be
\begin{eqnarray}
{Q}_1=-i \sqrt{2m}\int dx \psi^{\dagger} \phi
\end{eqnarray}

In order to write the supersymmetry algebra we define the Poisson brackets for the functions of the matter field as

\begin{eqnarray}
\lbrace F,G
\rbrace_{PB}=i\int dr \left( \frac{\delta F}{\delta \phi^\dagger(r)}
\frac{\delta G}{\delta \phi(r)}- \frac{\delta F}{\delta \phi(r)}
\frac{\delta G}{\delta \phi^\dagger(r)} - \frac{\delta^r F}{\delta
\psi^\dagger(r)} \frac{\delta^l G}{\delta \psi(r)}- \frac{\delta^r
F}{\delta \psi(r)} \frac{\delta^l G}{\delta \psi^\dagger(r)} \right)
\
\end{eqnarray}

where the subscripts $r$ and $l$ refer to right and left
derivatives and in particular we have

\begin{eqnarray}
\lbrace\phi(x_1,t),\phi^*(x_2,t)\rbrace=-i \delta(x_1-x_2)\;\;\;\,
\lbrace\psi(x_1,t),\psi^*(x_2,t)\rbrace=-i \delta(x_1-x_2)
\end{eqnarray}

Using the definition of Poisson bracket it is easy to get

\begin{eqnarray}
\lbrace Q_1,Q_1^\dagger \rbrace= -2 i m \int
dx \rho \equiv -2 i M \,\,.
\end{eqnarray}
Following the Ref.\cite{Van1} we can define a second generator

\begin{eqnarray}
R = -\frac{1}{\sqrt{2 m}} \int
dx \psi^\dagger \partial_x \phi \,\,.
\end{eqnarray}

which generate the free part of the Hamiltonian

\begin{eqnarray}
\lbrace
R,R^\dagger \rbrace=\frac{i}{2m} \int dx \;\;\;
(\phi^\dagger \partial_x^2 \phi + \psi^\dagger \partial_x^2 \psi) =-i H_{free}
\end{eqnarray}

Nevertheless, as we mentioned in the introduction, the previous works does not propose a supercharge be able to generate the full Hamiltonian of the model explored in Ref.\cite{Van1}, that is the Hamiltonian derived from the action (\ref{Ac5})

\begin{eqnarray}
H &=& \int d^2 x  \Big(
 \frac{1}{2m} ( \partial_x \phi )^\dagger  \partial_x \phi +\frac{1}{2m} ( \partial_x \psi )^\dagger  \partial_x \psi -
\nonumber \\  & & \frac{\gamma}{2m\kappa}( \rho_b -1)^2 -\frac{\gamma}{m\kappa} ( \rho_b -1)\rho_f   \Big)
\label{H}
\end{eqnarray}

In addition we know, from the Ref.\cite{my}, that the supersymmetry (\ref{t2}) is generated by the supercharge

\begin{eqnarray}
{Q_2^{(1)}}&=&- \frac{1}{\sqrt{2m}}\int dx \psi^\dagger(\partial_x+B_1)\phi
\nonumber \\
&=&- \frac{1}{\sqrt{2m}}
\int dx \psi^\dagger(x)(\partial_x+
\frac{\gamma}{2 \kappa} \int dz \epsilon(x-z) \rho(z)
)\phi(x)
 \end{eqnarray}
which generate a Hamiltonian only with potential terms quartic in the fields. Based in this idea and in the fact that the transformation (\ref{t3}) is a combination of the supersymmetry (\ref{t1}) and (\ref{t2}), it seems natural to define a supercharge $Q_2$, such that it is a linear combination of $Q_1$ and $Q_2^{(1)}$

\begin{eqnarray}
{Q_2}&=&- \frac{1}{\sqrt{2m}}\int dx \psi^\dagger(\partial_x+B_1)\phi + \frac{\sqrt{\gamma}}{\sqrt{\kappa m}} \int dx \psi^\dagger \phi
\label{Q2}
\end{eqnarray}

We will show that this charge generate the Hamiltonian (\ref{H}) . Using this charge we can calculate

\begin{eqnarray}
&&\left\{Q_2,Q_2^\dagger \right\} = \left\{ (K_1 + K_2),(K_1^\dagger + K_2^\dagger)\right\}= \nonumber \\
&&\left\{ K_1,K_1^\dagger \right\} + \left\{ K_1 ,K_2^\dagger\right\} + \left\{  K_2,K_1^\dagger \right\} + \left\{  K_2, K_2^\dagger\right\}
\end{eqnarray}

where

\begin{eqnarray}
K_1 = Q_2^{(1)} \;\;\;\,
K_2 = \frac{i\sqrt{\gamma}}{\sqrt{2 \kappa} m} Q_1
\end{eqnarray}

The first of the brackets was calculated in Ref.\cite{my} and its result is
\begin{eqnarray}
\left\{ K_1,K_1^\dagger \right\} &=& -\frac{i}{2m} \int  d^2 x  \Big(
\frac{1}{2m} ( \partial_x \phi )^\dagger  \partial_x \phi +\frac{1}{2m} ( \partial_x \psi )^\dagger  \partial_x \psi + \nonumber \\ & &
B_1^2 \rho - \frac{2}{\kappa} \rho_f \rho_b -\partial_x B_1 \rho_b + \partial_x B_1 \rho_f \Big)
\end{eqnarray}

which can be reduced, after using the definition of the $B_1$ and eliminating the constant term $\int  d^2 x B_1^2 \rho$, to

\begin{eqnarray}
\left\{ K_1,K_1^\dagger \right\} &=& -\frac{i}{2m} \int  d^2 x  \Big(
\frac{1}{2m} ( \partial_x \phi )^\dagger  \partial_x \phi +\frac{1}{2m} ( \partial_x \psi )^\dagger  \partial_x \psi - \nonumber \\ & &
\frac{2 \gamma}{\kappa} \rho_f \rho_b -\frac{\gamma}{\kappa}\rho_b^2 \Big)
\end{eqnarray}

The second bracket can be developed as follows

\begin{eqnarray}
\left\{ K_1,K_2^\dagger \right\} &=& -\frac{1}{m\sqrt{2\kappa}} \int dx_1 dx_2 \Big( \left\{\psi^\dagger(x_1),\psi (x_2)\right\}\phi^\dagger(x_2) (\partial_x+B_1 (x_1))\phi(x_1) +\nonumber \\ & &
\psi^\dagger(x_1)\left\{\partial_x \phi (x_1),\phi^\dagger(x_2)\right\}\psi (x_2) + \psi^\dagger(x_1)\left\{B_1(x_1),\phi^\dagger(x_2)\psi (x_2)\right\}\phi(x_1) + \nonumber \\ & &
\psi^\dagger(x_1)B_1(x_1)\left\{\phi(x_1),\phi^\dagger(x_2)\right\} \psi(x_2) \Big)
\end{eqnarray}

It can be easily checked that

\begin{eqnarray}
\left\{B_1(x_1),\phi^\dagger(x_2)\psi (x_2)\right\} =0
\end{eqnarray}

Thus

\begin{eqnarray}
\left\{ K_1,K_2^\dagger \right\} &=& \frac{i}{m\sqrt{2\kappa}} \int dx \Big( \phi^\dagger (x) (\partial_x + B_1 (x))\phi(x) + \nonumber \\ & &
\psi^\dagger (x) (\partial_x + B_1 (x))\psi(x) \Big)
\label {b2}
\end{eqnarray}

In similar form we have for the third bracket

\begin{eqnarray}
\left\{ K_2,K_1^\dagger \right\} &=& -\frac{i}{m\sqrt{2\kappa}} \int dx \Big( \phi^\dagger (x) (\partial_x - B_1 (x))\phi(x) +  \nonumber \\ & &
\psi^\dagger (x) (\partial_x - B_1 (x))\psi(x) \Big)
\label{b3}
\end{eqnarray}

From (\ref{b2}) and (\ref{b3}) we get

\begin{eqnarray}
\left\{ K_1,K_2^\dagger \right\} + \left\{ K_2,K_1^\dagger \right\} &=& \frac{i}{m\sqrt{2\kappa}} \int dx B_1 (x) \rho = \nonumber \\ & &
\frac{i}{2m\kappa} \int dx dz \epsilon(x-z)\rho(x)\rho(z)=0
\end{eqnarray}

The last bracket gives

\begin{eqnarray}
\left\{ K_2,K_2^\dagger \right\} = -\frac{i\gamma}{\kappa m} \int \rho dx
\end{eqnarray}

Then the full bracket takes the form

\begin{eqnarray}
\left\{ Q_2,Q_2^\dagger \right\} &=& -\frac{i}{2 m} \int  dx \Big( ( \partial_x \phi )^\dagger  \partial_x \phi + ( \partial_x \psi )^\dagger  \partial_x \psi -
\nonumber \\  & & \frac{2 \gamma}{\kappa} \rho_f \rho_b -\frac{\gamma}{\kappa}\rho_b^2 + \frac{2 \gamma}{\kappa} \rho   \Big) =-i H
\end{eqnarray}

The algebra is completed by the following bracket

\begin{eqnarray}
\left\{ Q_1,Q_2^\dagger \right\} &=& \left\{ Q_1,K_1^\dagger \right\} + \left\{ Q_1,K_2^\dagger \right\}
\end{eqnarray}

The first of this brackets can be calculated to give

\begin{eqnarray}
\left\{ Q_1,K_1^\dagger \right\} &=& -\frac{1}{2} \int d^2x
\Big(\phi^\dagger \partial_x\phi - \partial_x\phi^\dagger
\phi + \psi^\dagger\partial_x\psi -
\partial_x\psi^\dagger \psi \Big)
\end{eqnarray}

Where we have used that

\begin{eqnarray}
\int dx B_1 (x) \rho = 0
\end{eqnarray}

The second bracket is

\begin{eqnarray}
\left\{ Q_1,K_2^\dagger \right\} = - \sqrt{\frac{2\gamma}{\kappa}} \int dx \rho
\end{eqnarray}

Therefore

\begin{eqnarray}
\left\{ Q_1,Q_2^\dagger \right\} &=& -\frac{1}{2} \int d^2x
\Big(\phi^\dagger \partial_x\phi - \partial_x\phi^\dagger
\phi + \psi^\dagger\partial_x\psi -
\partial_x\psi^\dagger \psi \Big) - \nonumber \\  & &
\sqrt{\frac{2\gamma}{\kappa}} \int dx \rho
\end{eqnarray}

In order to show  that this expression may be identified with the linear momentum of the system we use the Noether's theorem. Let $\{\theta_c\} = \{\phi, \phi^\dagger, \psi, \psi^\dagger\}$ the set of the fields of the our system, where $c$ runs from 1 to 4. The theorem establish that if under a variation of the fields $\delta \theta_c$, the variation of the Lagrangian density is a surface term, $\delta {\cal{L}}= \partial_\mu X^\mu$, then exist a conserved current associated with such variation of the fields. The Noether current, assuming the summation convention over the index c, is

\begin{eqnarray}
j^{\mu} = \frac{\partial \cal{L}}{\partial(\partial_\mu \theta_c)}\delta \theta_c - X^\mu
\label{Noe}
\end{eqnarray}

We are interested on the zero component of this current associated to the transformations

\begin{eqnarray}
\{\delta\theta_c\} = \{\partial_x\phi, \partial_x\phi^\dagger, \partial_x\psi, \partial_x\psi^\dagger\}
\end{eqnarray}

where

\begin{eqnarray}
X^0 &=& -i \gamma\sqrt{\frac{2\gamma}{\kappa}}  \rho \;, \;\;\;
\;\;\; X^1= 0\;,
\label{}
\end{eqnarray}

Note that, from the Noether theorem, the only restriction for  $X^\mu$ is that  $\partial_\mu X^\mu$ be a surface term.
Using (\ref{Noe}), the linear momentum is

\begin{eqnarray}
P =\int j^{0} dx = -i \gamma \{ Q_1, Q_2^\dagger\}
\end{eqnarray}

The energy-momentum-tensor, then may be defined as

\begin{eqnarray}
T^\mu_\nu = \frac{\partial \cal{L}}{\partial(\partial_\mu \theta_c)} \partial_\nu \theta_c - {\cal{L}} \delta^\mu_\nu + i \gamma\sqrt{\frac{2\gamma}{\kappa}}  \rho \tilde{\delta}^\mu_\nu
\end{eqnarray}

where $\tilde{\delta}^\mu_\nu =1$ if $\mu \not= \nu$ and $\tilde{\delta}^\mu_\nu =0$ if $\mu = \nu$.

Finally it is easy to check that the remaining brackets are zero

\begin{eqnarray}
\left\{ Q_\alpha,Q_\beta \right\} = \left\{ Q_\alpha^\dagger,Q_\beta^\dagger \right\} =0
\end{eqnarray}

\section{The soliton solution}

Consider now the derivation of the self-dual equations. As discussed in Ref.\cite{OH} the field $B$ plays an important role in the derivation of self-dual equations. Indeed the expression (\ref{gauss}) of $B$  involves the existence of a novel soliton. The action (\ref{Ac4}) may be easily reexpressed as

\begin{eqnarray}
 S& =& \int d^2 x \Big( i\gamma \lbrace   \phi^\dagger \partial_t \phi +
\psi^\dagger \partial_t \psi \rbrace - \frac{1}{2m} | (\partial_x + \zeta
B)\phi|^2  - \frac{1}{2m} |( \partial_x + \zeta B)\psi|^2
 - \nonumber \\
& &\frac{\zeta}{2m} \partial_x B \rho + \lambda^{'}_1 (\rho_b - 1)^2 +
\lambda^{'}_{2}(\rho_b - 1)\rho_f \Big)\; ,
\end{eqnarray}

where $\zeta =\pm 1$. Using the Gauss law (\ref{G2}) and after a bit of algebra we have

\begin{eqnarray}
 S& =& \int d^2 x \Big( i\gamma \lbrace   \phi^\dagger \partial_t \phi +
\psi^\dagger \partial_t \psi \rbrace - \frac{1}{2m} | (\partial_x + \zeta
B)\phi|^2  - \frac{1}{2m} |( \partial_x + \zeta B)\psi|^2
 + \nonumber \\
& & (\lambda^{'}_1 - \frac{\zeta \gamma}{2m\kappa}) (\rho_b - 1)^2 +
(\lambda^{'}_{2} -\frac{\zeta \gamma}{m\kappa})(\rho_b - 1)\rho_f -\frac{\zeta}{2m}\partial_x B \Big)\; ,
\end{eqnarray}

which leads, in the static field configuration, to the Hamiltonian of the form,

\begin{eqnarray}
 H& =& \int d^2 x \Big(  \frac{1}{2m} | (\partial_x + \zeta
B)\phi|^2  + \frac{1}{2m} |( \partial_x + \zeta B)\psi|^2
- \nonumber \\
& & (\lambda^{'}_1 - \frac{\zeta \gamma}{2m\kappa}) (\rho_b - 1)^2 -
(\lambda^{'}_{2} -\frac{\zeta \gamma}{m\kappa})(\rho_b - 1)\rho_f +\frac{\zeta}{2m}\partial_x B \Big)
\end{eqnarray}

We can choose $\lambda^{'}_1 = \frac{\zeta \gamma}{2m\kappa}$ and $\lambda^{'}_{2} = \frac{\zeta \gamma}{m\kappa}$ so that our Hamiltonian becomes

\begin{eqnarray}
 H& =& \int d^2 x \Big(  \frac{1}{2m} | (\partial_x + \zeta
B)\phi|^2  + \frac{1}{2m} |( \partial_x + \zeta B)\psi|^2
+\frac{\zeta}{2m}\partial_x B \Big)
\end{eqnarray}

The last integral vanish since $B$ must be zero in de boundary. Then, at the minimum of the energy configurations, the self-dual equations are satisfied

\begin{eqnarray}
& &(\partial_x +\zeta
B)\phi=0 \\
& &
(\partial_x +\zeta
B)\psi=0
\end{eqnarray}

Notice that for the particular choice $\zeta=1$, we recover the supersymmetric case. This equations can be explicitly written by using the equation (\ref{gauss})

\begin{eqnarray}
 & & \partial_x\phi(x) + \frac{\zeta\gamma}{2 \kappa} \Big(\int dz \epsilon(x-z)\rho(z)\phi(x) -\int dz \epsilon(x-z)\phi(x) \Big)=0 \\
& &
\partial_x \psi + \frac{\zeta\gamma}{2 \kappa} \Big(\int dz \epsilon(x-z)\rho(z)\psi(x) -\int dz \epsilon(x-z)\psi(x) \Big)=0
\end{eqnarray}

which present an additional linear term from those found in Ref.\cite{my}. When $\psi$ is set to zero, the above set of equations reduces to

\begin{eqnarray}
 & & \partial_x\phi(x) + \frac{\zeta\gamma}{2 \kappa} \int dz \epsilon(x-z)(\rho_b(z)-1)\phi(x) =0
\end{eqnarray}

Assuming a solution of the form $\phi=\sqrt{\rho_b}$, we arrive to

\begin{eqnarray}
 & & \frac{1}{2}\partial_x (\log \rho_b (x)) + \frac{\zeta\gamma}{2 \kappa} \int dz \epsilon(x-z)(\rho_b(z)-1) =0
\end{eqnarray}

Differentiating with respect to $x$, we get the following one-dimensional Liouville type equation

\begin{eqnarray}
 & & \frac{1}{2}\partial_x^2 (\log \rho_b (x)) + \frac{\zeta\gamma}{2 \kappa}(\rho_b(x)-1) =0
 \label{Lio}
\end{eqnarray}

We propose as the solution of this equation the following series

\begin{eqnarray}
 \rho = 1 + \sum_{n=1}^\infty a_n \rm{sech}^n (b x)
 \label{rho}
\end{eqnarray}

Here $a_n$ are the real coefficients of series, $b$ is a real constant and we have renamed $\rho$ as $\rho_b (x)$.
In order to check that this is really a solution we rewrite Eq.(\ref{Lio} ) as follows

\begin{eqnarray}
 & & -(\partial_x \rho)^2 + (\partial_x^2 \rho)\rho + \upsilon \rho^2(\rho -1) =0
 \label{Lio2}
\end{eqnarray}

where $\upsilon = \frac{\zeta\gamma}{ \kappa}$.
When the series (\ref{rho}) is introduced into  Eq.(\ref{Lio2}) we obtain

\begin{eqnarray}
& &\sum_{n,m=1}^\infty \Big[ -a_n a_m m n b^2 + n^2a_n a_m b^2 + 2\upsilon a_n a_m \Big]  \rm{sech}^{n+m}(bx) +
\nonumber \\
& &\sum_{n,m=1}^\infty \Big[ a_n a_m n m b^2 - n^2a_n a_m b^2 - n a_n a_m b^2 \Big] \rm{sech}^{n+m+2} (bx) -
\nonumber \\
& &\sum_{n=1}^\infty (n^2 +n)a_n b^2 \rm{sech}^{2+n} (bx) + \sum_{n=1}^\infty \Big[ n^2 a_n b^2 + \upsilon a_n \Big] \rm{sech}^n (bx) +
\nonumber \\
& &\upsilon \sum_{n,i,m=1}^\infty a_n a_m a_i \rm{sech}^{n+m+i}(bx) =0
\end{eqnarray}

where for arriving to this expression we have used the relation

\begin{eqnarray}
\tanh^2 (bx) =1 - \rm{sech}^2 (bx)
\end{eqnarray}

So we have an expansion of powers of $\rm{sech}(bx)$ which must be equal to zero. This implies that the coefficient of each power must vanish separately. From the coefficient of $\rm{sech}(bx)$ we obtain

\begin{eqnarray}
b^2=-\upsilon
\end{eqnarray}

whereas from the coefficients of  $\rm{sech}^2(bx)$ and $\rm{sech}^3(bx)$ we have

\begin{eqnarray}
a_2=\frac{2}{3}a_1^2
\end{eqnarray}

and

\begin{eqnarray}
a_3 = \frac{3a_1^3 + 2a_1}{8}
\end{eqnarray}

The method can be continued  in order to determine the  rest of the coefficients.

\section{Conclusion}

In this article we have studied a (1+1)-dimensional model introduced in the description of the supersymmetric-ultracold gases.
This model and its supersymmetries was previously studied in Ref.\cite{Van2,Van1,my}. However the problem of finding the supersymmetry algebra that generate the full theory was unresolved. In this note we started by extending supersymmetrically a model proposed by Manton, and related it to the theory of ultracold gases. Then, the correct supercharge that generate the full theory were identified and their algebra was constructed. In addition the solitonic structure was analyzed and novel solitons are found.

\vspace{2cm}
{\bf Acknowledgements}
I'm extremely grateful to Fidel Schaposnik for spent patiently many hours listening and explaining.

\end{document}